\documentclass[aps,prl,preprint,groupedaddress]{revtex4-1}
\usepackage{amsmath}
\usepackage{amssymb}
\usepackage{graphicx}
\def\equ#1{(\ref{#1})}
\def\be#1{\begin{equation}\label{#1}}
           
       \def\ee{\end{equation}}

\begin{document}
 \title{Analytical solutions for fermions on a thick brane with a piecewise and smooth
warp factor}

\author{R. R. Landim${}^a$}
\email{rrlandim@gmail.com}
\author{M. O. Tahim${}^{b,}$}
\email{makarius.tahim@gmail.com}
\author{G. Alencar${}^a$}
\email{geova@fisica.ufc.br}
\author{R.N. Costa Filho${}^a$}
\email{rai@fisica.ufc.br}
\affiliation{${}^a$Departamento de F\'{\i}sica, Universidade Federal
do Cear\'a, 60451-970 Fortaleza, Cear\'a, Brazil}
\affiliation{${}^b$Universidade Estadual do Cear\'a, Faculdade de Educa\c c\~ao, Ci\^encias e Letras do Sert\~ao Central - 
R. Epit\'acio Pessoa, 2554, 63.900-000  Quixad\'{a}, Cear\'{a},  Brazil.}
\date{\today}
\begin{abstract}
In this paper we study analytical solutions for fermion localization in Randall-Sundrum (RS) models. We 
show that there exist special couplings between scalar fields and fermions giving us discrete massive localizable modes. 
Besides this we obtain resonances in some models by analytical methods.
\end{abstract}

\maketitle

\section{Introduction}

The study of gravity resonances in Randall-Sundrum models \cite{Randall:1999vf,Randall:1999ee}  is of great importance because, since stable states for most fields do not exist, it points to the possibility of finding unstable massive modes in the membrane. Much of these studies used smooth versions of the background with the RS warp factor recovered for large 
$z$\cite{Bazeia:2005hu,Liu:2009ve,Zhao:2009ja,Liang:2009zzf,Alencar:2010hs,Landim:2010pq,Landim:2011ki,Landim:2011ts}. The emergence of resonances is an interesting phenomenological possibility that has been 
approached mainly numerically. However, some examples have been studied where an 
integrable solution was found for bosonic fields 
\cite{Cvetic:2008gu,Alencar:2012en}.  In order to obtain these solutions a specific form 
of the background was chosen such that a simple potential was obtained for an 
associated Schr\"odinger equation. That leads to the question: what kind of 
background allows analytical solutions? The answer to this question was 
given in \cite{Landim:2013dja} where the authors found an associated equation 
leading to the correct background. That was possible by the introduction of an 
arbitrary function describing the region near the (thick) brane. This arbitrary 
function makes the connection with the effective potential of the Schr\"odinger 
like equation. With this, the backgrounds used in the literature 
\cite{Cvetic:2008gu,Alencar:2012en,Li:2010dy} are just particular solutions of 
this equation. Besides, many others solutions were found.  In particular, the 
fermion case was studied in ref. \cite{Li:2010dy} but the solutions were found 
only numerically. It happens that the fermion case introduces additional 
complications because of the coupling with the scalar field. Anyway, it is 
possible to circumvent this and generalize the idea used for the gravity case: 
here we find a few examples of analytical solutions. Therefore, the complete 
structure of resonances associated with each of these new solutions are also 
obtained. The way chosen to describe the background of the thick brane gives us a better
understanding of its structure and can lead to new possibilities. 

This letter is organized as follows: in the first section we give a brief review of spin-1/2 fermion background. In the second section we describe fermion in a profile with piecewise warp factor. The third section is devoted to study fermion in a background with smooth warp factor. Finally, we present the conclusions and perspectives.

\section{The fermion setup}

We regard here the problem of studying fermions in a Randall-Sundrum like scenario. Usually it is 
considered the coupling between fermions and the gravitational fields alongside 
a coupling between fermions and a scalar field $\phi$, the stuff the membrane 
is made of. Indeed, the membrane is regarded as a domain wall embedded in $D=5$. 
Studies of fermion localization in this context reveals a very natural mechanism 
of producing chiral fields in $D=4$ 
\cite{Kehagias:2000au,Rubakov:1983bb,Liu:2013kxz,Chumbes:2010xg,Castro:2010uj,
Correa:2010zg,Castro:2010au,CastilloFelisola:2010xh,Liang:2009zzf,
Zhao:2009ja,Liu:2009dw,Liu:2009ve,Melfo:2006hh}. The new 
ingredient here is that, if the warp factor is defined in a piecewise way, so 
must be the scalar field $\phi$. The usual coupling is the Yukawa one that will be generalized to allow a more complete analysis.

First, we  give a brief review of the fermion background.  Let us consider a single scalar field thick brane $\phi$ 
with a  potential $V(\phi)$ in (4+1)-dimensional curved space-time with cosmological constant $\Lambda$. The action describing this model 
is given by
\begin{eqnarray}
 S = \int d^{5}x \sqrt{-g}
  \left[ \frac{1}{2\kappa_{5}^{2}}(R -2\Lambda)-\frac{1}{2}g^{MN}\partial_{M}\phi\partial_{N}\phi
  -V(\phi) \right]\,,~~
 \label{scalarAction}
\end{eqnarray}
where $R$ is the scalar curvature, $g = \det[g_{MN}]$, $M,N = 0,1,2,3,4$, and
$\kappa_{5}^{2}=8\pi G_{5}$ with $G_{5}$ the five-dimensional Newtonian
gravitational constant. We will use the metric in the conformal coordinates and the line-element is assumed to be
\begin{eqnarray}
 ds^{2} =g_{MN}dx^{M}dx^{N}
        =e^{2A(z)}(\eta_{\mu\nu}dx^{\mu}dx^{\nu}+dz^{2}), \label{metric}
\end{eqnarray}
where $e^{2A(z)}$ is the warp factor. The metric tensor $\eta_{\mu\nu}$ presents
signature in the form $(-,+,+,+,+)$, and $z$ stands for the
coordinate of the extra spatial dimension. The scalar field is considered to be a function of $z$ only: $\phi = \phi(z)$.  From the metric ansatz \equ{metric} the Einstein tensor is given by
\begin{eqnarray}
G_{\mu\nu}&=&(3A^{\prime2}+3A^{\prime\prime})\eta_{\mu\nu}\,,\quad G_{44}=6A'^2\,,   \label{Einstein_tensor} \\
G_{4\mu}&=&0\,,
\end{eqnarray}
where the prime denotes derivative with respect to the $z$ coordinate.

The Euler-Lagrange equations for the action (\ref{scalarAction})
with the metric ansatz (\ref{metric}) after some manipulations are given by:
\begin{eqnarray}
 \kappa_5^2\phi^{\prime2} &=& 3(A^{\prime2}-A^{\prime\prime})\,,   \label{2ndOrderEqs1} \\
 V(\phi) &=& -\frac{1}{2\kappa_5^2}(9A^{\prime2}+3A^{\prime\prime})e^{-2A}-\frac{\Lambda}{\kappa_5}\,,  \label{2ndOrderEqs2} \\
 \frac{d V(\phi)}{d\phi} &=& (3A^{\prime}\phi^{\prime}+\phi^{\prime\prime})e^{-2A} \,. \label{PotentialV1}
\end{eqnarray}

If the $Z_2$ symmetry is imposed, namely $A(z)=A(-z)$,  we can see
from Eqs. \equ{2ndOrderEqs1} and \equ{2ndOrderEqs2}  that in the limit $|z|\rightarrow \infty$ with $\phi^{\prime}(z)\rightarrow0$ and  $V(\phi)\rightarrow0$ we have  $A(z)\rightarrow -\ln(k_0 |z|+\beta)$ and $\Lambda=-6k_0^2.$ In this case the bulk is asymptotically AdS. It is worthwhile to mention that the condition $\lim_{|z|\rightarrow\infty}\phi^{\prime}(z)\rightarrow0$ can be satisfied by  kink solution $(\lim_{z\rightarrow \pm\infty}\phi(z)=\pm c_0,c_0>0)$. We shall restrict the bulk  asymptotically AdS  and kink solutions (with $\phi(z)=-\phi(-z)$) through this paper. We also make $\kappa_5=1$ for simplicity.

In this background, the Dirac fermionic action for a spin 1/2 field coupled to the scalar field $\phi$ is

\begin{eqnarray}
 S_{1/2} =\int d^{5}x\sqrt{-g}\left[\overline{\Psi}\Gamma^{M}D_{M}\Psi
    -\eta\overline{\Psi}F(\phi)\Psi\right],
    \label{ffa}
\end{eqnarray}
where $D_M=\partial_M+\omega_M$, $\omega_M$ being the spin connection. The $\Gamma^M=e_{N}^{~~M}\gamma^N$ are the Dirac matrices in the five dimensional curved space-time 
and $e_{N}^{~M}$ are the vielbeins: $e_{A}^{~~M}e_{B}^{~~N}\eta^{AB}=g^{MN}$. 

The Dirac equation in five-dimensional space-time can be derived from the 
action\equ{ffa}
\begin{eqnarray}
\left[\gamma^{\mu}\partial_{\mu}+\gamma^{5}\left(\partial_{z}+2\partial_{z}
A\right)-\eta
e^{A}F(\phi)\right]\Psi(x,z)=0\,,
 \label{DiracEq1}
\end{eqnarray}
where $\gamma^{\mu}\partial_{\mu}$ is the four-dimensional Dirac
operator on the brane.

Due to the presence of $\gamma^5$ in the Eq. \equ{DiracEq1}, is convenient to 
make the chiral decomposition of the fermions in Kaluza-Klein (KK) 
modes\cite{Li:2010dy}:
\begin{eqnarray}
\Psi(x,z)=e^{-2A}\sum_{n}\left[\psi_{\text{L}n}(x)\psi_{-}^n(z)
+\psi_{\text{R}n}(x)\psi_{+}^n(z)\right]\,,\nonumber\\
\label{the general chiral decomposition}
\end{eqnarray}
where $\psi_{\text{L}n}(x)=-\gamma^{5}\psi_{\text{L}n}(x)$ and
$\psi_{\text{R}n}(x)=\gamma^{5}\psi_{\text{R}n}(x)$ are the left-handed and 
right-handed components of the Dirac fermion fields on the brane, respectively. They satisfy 
the four-dimensional massive Dirac equations in the form of
$\gamma^{\mu}\partial_{\mu}\psi_{\text{L}n}(x)=m_{n}\psi_{\text{R}n}(x)$ and
$\gamma^{\mu}\partial_{\mu}\psi_{\text{R}n}(x)=m_{n}\psi_{\text{L}n}(x)$. The 
KK modes $\psi_{\pm}^n(z)$ obeys the coupled equations

\be{CP}
\frac{d\psi_{\pm}}{dz}\mp\eta e^{A(z)}F(\phi(z))\psi_{\pm}=\mp m\psi_{\mp}.
\ee
From these we obtain the 
Schr\"odinger like equation for each mode 
\begin{equation}\label{Sch}
-\psi''_{\pm}(z)+U_{\pm}(z)\psi(z)_{\pm}=m^2\psi_{\pm}(z),
\end{equation}
where the  effective potential $U(z)$ depends on the warp factor as bellow
\begin{equation}\label{U(A)}
U_{\pm}(z)=(\eta F(\phi(z))e^{A(z)})^2{\pm}\frac{d}{dz}(\eta 
F(\phi(z))e^{A(z)}).
\end{equation}

The study of the massive modes must be done more carefully. This happens because when
the potential in \equ{Sch} has the asymptotic behaviour $\lim_{|z|\rightarrow\infty} U_{\pm}(z)=0$,   $\psi_{\pm}(z)$ are linear combination of plane waves and therefore not normalizable.

For massive modes, like in non-relativistic quantum mechanics, we will use the transmission coefficient to study possible resonances\cite{Landim:2011ki}.
Usually, the transmission coefficient represents the 
probability flux of the transmitted wave relative to  the incident wave. In our case, the transmission coefficient represents the  probability 
of a fermionic massive mode to pass  through the brane.
The idea of 
the existence of a resonant mode is that for a given mass the transmission coefficient has a peak at this mass value. 
In our study of resonances we limit the solutions of \equ{Sch} that have asymptotic planes wave form.

\section{The piecewise case}

In this section the warp factor $A(z)$ will be defined 
in two regions, $|z|\leq d$ and $|z|\geq d$:
\begin{eqnarray}\label{A}
A(z)=
\begin{cases}
-\ln(f(z)), &|z|\leq d;\\
~\\
-\ln[(k_0(|z|+\beta)], & |z|\geq d,
\end{cases}
\end{eqnarray}
where $d$, beta are positive constants and $f(z)$ is called associated 
function with $f(z)>0$ in $|z|\leq d$, $k_0=\sqrt{-\Lambda/6}$. This is 
enough to establish how the membrane curves the spacetime with  the bulk being asymptotically AdS\cite{Alencar:2012en,Cvetic:2008gu}.

In order to get the desired smooth versions, the functions $A$,  $A'$ and $A''$ must be continuous to give $\phi$ and $\phi'$ 
continuous at
$|z|=d$. That restricts the form of $f(z)$
in the membrane. For the gravitational field the choice of \cite{Cvetic:2008gu} was $A(z)=\frac{2}{3}\ln cos(\sqrt{V_{0}}|z|)$ giving the 
effective potential $U=-V_{0}$ for $|z|\leq d$. That allows a detailed analytical study of the model. 
Recently, this result was found as a particular one belonging to a wider class of new exact solutions for the gravitational
field \cite{Landim:2013dja}. The purpose here is to present a new way to generalize this idea in order to obtain 
analytical solutions for the fermion field as well. 

We focus our attention in right fermions. The analysis is similar for left fermions. In order to implement the boundary conditions, we restrict to even
functions $f(z)$ in \equ{A}, guaranteeing that
the boundary conditions are satisfied in both edges of the brane,
$z=d$ and $z=-d$.  The condition $A(0)=0$ implies that
$f(0)=1$. Since $f(z)$ is an even function we also have $A'(0)=0$. 
Using the equations \equ{2ndOrderEqs1} and \equ{2ndOrderEqs2} we obtain
\begin{eqnarray}\label{fil}
\phi'(z)^ 2=
\begin{cases}
3\frac{f''(z)}{f(z)}, &|z|\leq d,\\
~\\
0, & |z|\geq d,
\end{cases}
\end{eqnarray}
and 

\begin{equation}
 V(\phi(z))=\frac{3}{2}f''(z)f(z)-6f'(z)^2+6k_0^2.\label{vphipiece}
\end{equation}

From  Eq. (\ref{CP}) we can see  that the massless right modes  obey the condition
\begin{equation}\label{cond0}
\frac{f_0'(z)}{f_0(z)}=\eta e^{A(z)}F(\phi(z)).
\end{equation}
With this condition, the potential of the schrodinger-like equation becomes
\begin{equation}\label{Ueqf1}
U(z)=\frac{f_{0}''(z)}{f_{0}(z)}.
\end{equation}
If we choose $f_0(z)$ such that \equ{Sch} is analytically solved  we can obtain analytically the resonances.

In order to implement the boundary conditions, we are restricted to even functions $f_0(z)=f_0(-z)$ with $U(z)=U(-z)$ in \equ{Ueqf1}. This guarantees that the boundary condition in $z=d$ is satisfied also in $z=-d$. 
Our method consist in finding $U(z)$ and  $F(\phi(z))$ that have analytical 
solution for Eq. (\ref{Sch}). As we can see from \equ{cond0}, $F(\phi(z))$ is an odd function in $z$.  From Eq. \equ{fil}, 
$\phi(z)=c_0$ for $z\ge d$. We also have two regions 
  for $f_0(z)$:
  \begin{eqnarray}\label{f_0}
f_0(z)=
\begin{cases}
g_0(z), &|z|\leq d,\\
~\\
(|z|+\beta)^{\gamma}, & |z|\geq d,
\end{cases}
\end{eqnarray}
where $\gamma=\eta F(c_0)/k_0$ and $g_0(z)$ being an even function. In order to obtain 
a localizable zero mode, $\gamma<-1/2$. Now the potential in \equ{Ueqf1} becomes
\begin{eqnarray}
U(z)=
\begin{cases}
\frac{g''_0(z)}{g_0(z)}, &|z|\leq d,\\
~\\
\frac{\gamma(\gamma-1)}{(|z|+\beta)^2}, & |z|\geq d.
\end{cases}
\end{eqnarray}
The solution of \equ{Sch} for $|z|\geq d$ is a linear combination of Bessel 
functions of order $|\gamma|+1/2$.
With the above considerations and by imposing continuity of $A(z)$, $A'(z)$, 
$A''(z)$, $f_0(z)$ and $f'_0(z)$ at $z=\pm d$  we obtain 
\begin{eqnarray}
 f(d)=k_0(d+\beta), \label{fd}\\
 f'(d)=k_0,\label{fdl}\\
 f''(d)=0,\label{fdll}
\end{eqnarray}
\begin{eqnarray}
 g_0(d)=(d+\beta)^\gamma, \label{g0d}\\
 g_0'(d)=\gamma(d+\beta)^{\gamma-1}\label{g0dl},
\end{eqnarray}
with the conditions $f(d)>0$. 

The solution for right fermion of \equ{Sch} will be given by
\begin{eqnarray}\label{psim}
\psi_m(z)=
\begin{cases}
 E_m(z)+C_mF_m(z), & z\leq -d,\\
 A_mg_m(z)+B_mh_m(z), &|z|\leq d,\\
D_m F_m(z), & z\geq d,  
\end{cases}
\end{eqnarray}
where $E_m(z)=\sqrt{\frac{\pi u}{2}}H_{\nu}^{(2)}(u),F_m(z)=\sqrt{\frac{\pi u}{2}}H_{\nu}^{(1)}(u)$, with $u=m(|z|+\beta)$,
$\nu=(|\gamma|+1/2)$ and $g_m(z)(h_m(z))$ are the even(odd) functions. The functions $E_m(z)$ and $F_m(z)$ have asymptotic behaviour $e^{imz}$.
In order to analyze the resonances, we fixed the coefficient of the incoming
wave  to one. 

\begin{figure}
 \centering
 \includegraphics[scale=0.5]{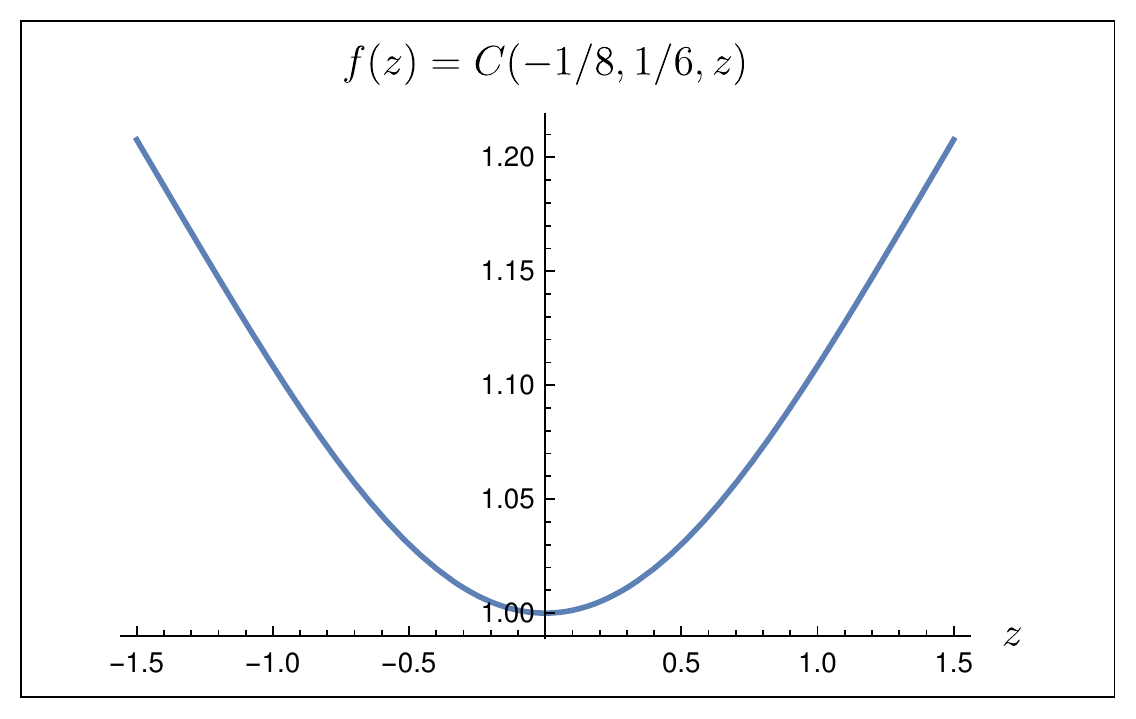}
 \caption{The Mathieu cosine function $C(-1/8,1/16,z)$.}
 \label{fig1}
\end{figure}

Using the  fact that $g_m(z),E_m(z),F_m(z)$ are even and $h_m(z)$ are odd and 
the continuity of the $\psi_m(z)$, $\psi'_m(z)$ at $|z|=d$ we
finally obtain the transmission coefficient

\begin{equation}\label{analictt}
T(m)=\frac{m^2|W(g_m,h_m)(d)|^2}{|W(F_m,g_m)(d)W(F_m,h_m)(d)|^2},
\end{equation}
where $W(f,g)(z)=f(z)g'(z)-f'(z)g(z)$ is the Wronskian.
As applications of the method, consider the scalar field in $|z|\leq d$ given by 
$\phi(z)=a\sqrt{3}\sin(z)$. From
Eq. \equ{fil} we obtain
\begin{equation}\label{pot-osc}
-f''(z)+a^2\cos^2(z)f(z)=0.
\end{equation}
The even solution of \equ{pot-osc} is given by the Mathieu cosine functions\cite{mclanh}:
\begin{equation}
f(z)=C\left(-\frac{a^2}{2},\frac{a^2}{4},
z\right)\; \label{funf},
\end{equation}
with $f(z)$ real and $f(0)=1$.
From Eqs. \equ{fd}, \equ{fdl} and \equ{fdll},  we obtain $d=\pi/2$, 
$k_0=0.65447$, $\beta=1.02593$ for $a=1/2$. 
The scalar potential can be written as
\begin{eqnarray}
V(\phi)= \frac{3}{8}(1-\frac{4\phi^2}{3})C^2\left(-\frac{1}{8},\frac{1}{16},\arcsin(2\phi/\sqrt{3})\right)\nonumber\\
-6C'^2\left(-\frac{1}{8},\frac{1}{16},\arcsin(2\phi/\sqrt{3}),
\right)+6k_0^2
\end{eqnarray}
where $C'(n,m,x)$ is the derivative of $C(n,m,x)$ with respect do $x$. The function $C(a,b,x)$ is obtained from the real part of $MathieuC(a,b,x)$ divided by the 
real part of $MathieuC(a,b,0)$, where $MathieuC(a,b,x)$ is defined in the Mathematica software. For completeness, we show the plot of $f(z)=C(-1/8,1/16,z)$ in Fig. \ref{fig1}.

\begin{figure}
\centering
 \includegraphics[scale=0.5]{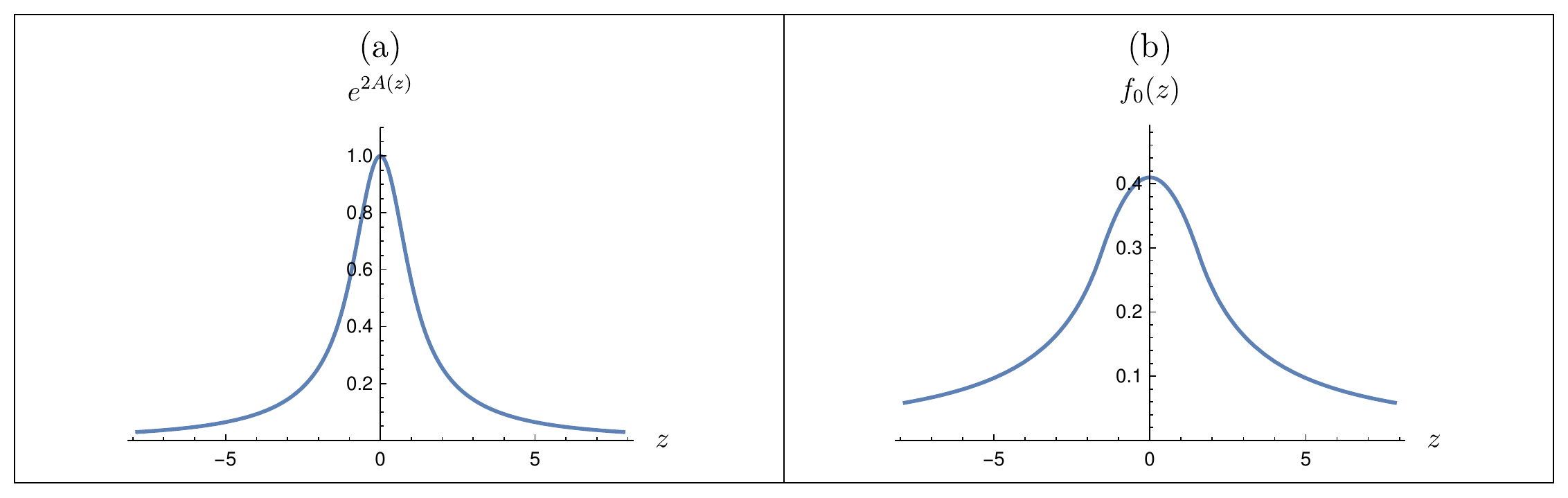}
 \caption{(a) The exponential of two times the warp factor. (b) The zero mode 
of right fermion for $a=1/2$ and $V_0=1/4$.}
 \label{fig2}
\end{figure}

For $g_0(z)=A_0\cos(\sqrt{V_0}z), V_0>0$ in \equ{f_0} we obtain $U(z)=-V_0$, 
$|z|\leq 
\pi/2$. From Eqs. \equ{g0d} and \equ{g0dl} we have
\begin{eqnarray}
&&\gamma=-(\frac{\pi}{2}+\beta)\sqrt{V_0}\mbox{tan}(\sqrt{V_0}\frac{\pi}{2}),\\
&&A_0=(\frac{\pi}{2}+\beta)^\gamma/\cos(\sqrt{V_0}\frac{\pi}{2}).
\end{eqnarray}
Since $\gamma<-1/2$, we have some restrictions for the values of $V_0$. We also 
have that $\cos(\sqrt{V_0}\frac{\pi}{2})\ne0$ 
and $\sin(\sqrt{V_0}\frac{\pi}{2})\ne0$. This implies that $\sqrt{V_0}$ cannot 
be an integer. For $V_0=1/4$ we have $\gamma=-1.29836$ and $A_0=0.409675$. We show in Fig. \ref{fig2}, the exponential of two times the 
warp factor  and the massless right fermion wave function for $V_0=1/4$.

Now using the equation \equ{cond0}, we show below the scalar fermion coupling
\begin{equation}\label{fphi1}
 \eta F(\phi)=\frac{-2A_0}{\sqrt{3}+\sqrt{3-4\phi^2}}C\left(-\frac{1}{8},\frac{1}{16},\arcsin(2\phi/\sqrt{3})\right).
\end{equation}

\begin{figure}
 \centering
 \includegraphics[scale=0.5]{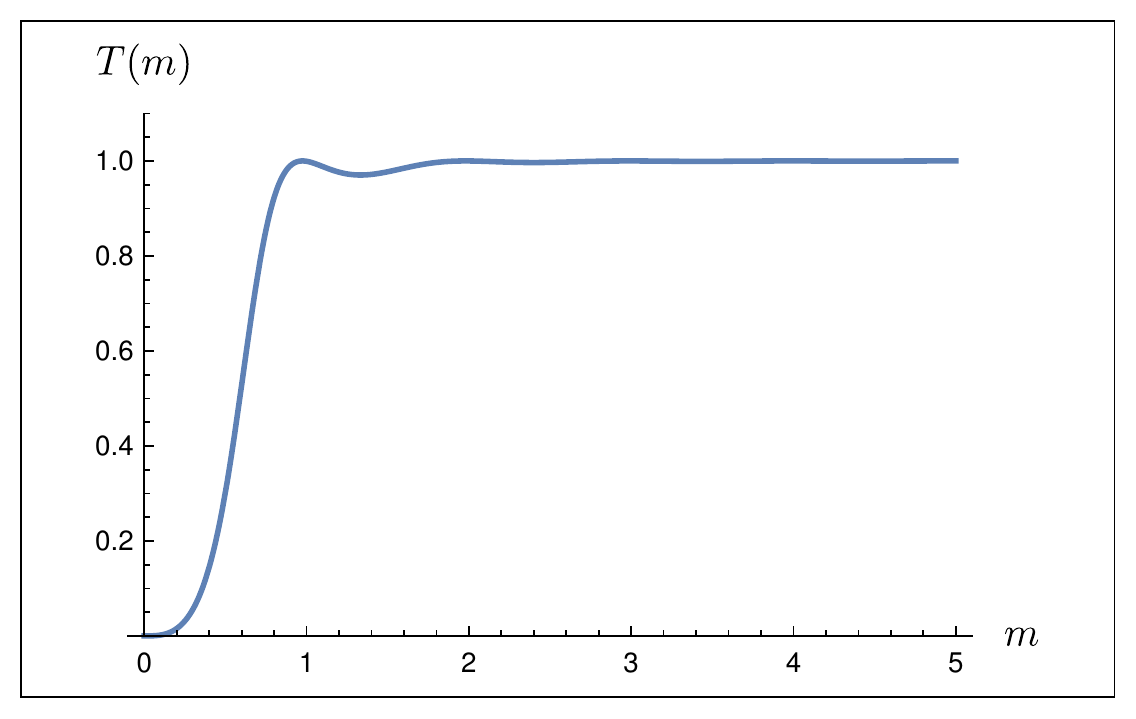}
 \caption{The transmission coefficient 
of right fermion for $a=1/2$ and $V_0=1/4$.}
 \label{fig3}
\end{figure}

The 
solution of \equ{Sch} for $|z|\leq d$ is a linear combination of 
$g_m(z)=\cos(\sqrt{m^2+V_0}z)$ and $h_m(z)=\sin(\sqrt{m^2+V_0}z)$. Now using the Eq. \equ{analictt}, we show in Fig. \ref{fig3} the transmission coefficient for the right fermion. 
As physical result, we have a massive resonant mode near $m=0.9$.

\section{The smooth case}

In this section we consider a smooth warp factor given by
\be{A3}
A(z)=-\frac{1}{2}\ln (k_0^2 z^2+1), \quad -6k_0^2=\Lambda.
\ee
This type of warp factor was used in \cite{Du:2013bx} to obtain resonances for Kalb-Ramond fields. From Eqs. \equ{2ndOrderEqs1} and \equ{2ndOrderEqs2} we
obtain

\begin{eqnarray}
 \phi(z)=\sqrt{3}\mbox{arctan}(k_0z),\\
 V(\phi)=\frac{15k_0^2}{2}\cos^2(\frac{\phi}{\sqrt{3}}).
\end{eqnarray}

\begin{figure}[ht]
 \centerline{\includegraphics[scale=0.5]{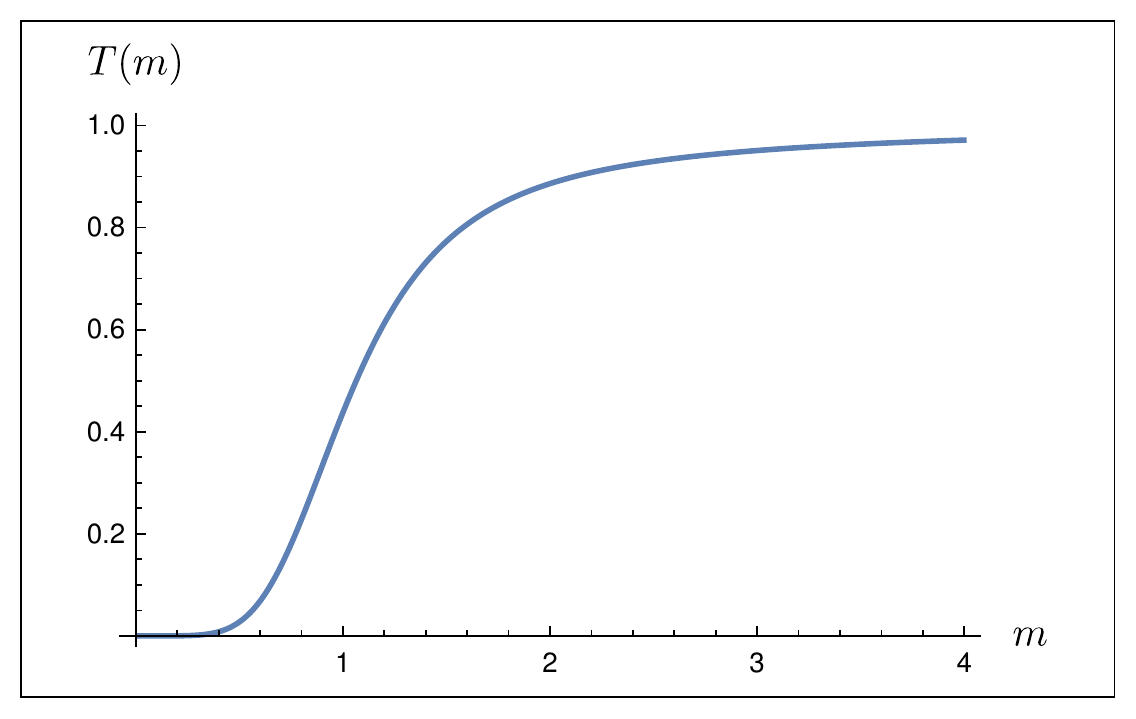}}
 \caption{The transmission coefficient 
of right fermion for Sine-Gordon coupling, with $k_0=1, \eta=-1$.}
 \label{fig4}
\end{figure}

We consider for now, the Sine-Gordon-like fermion coupling: $F(\phi)=\sin(\phi/\sqrt{3})$. With this we obtain the potential
\begin{equation}
 U(z)=\frac{k_0 \eta \left(-k_0^2 z^2+k_0 \eta z^2+1\right)}{\left(k_0^2 z^2+1\right)^2}.
\end{equation}
Making the substitution $\psi(z)=\Phi(z)(1+k_0^2 z^2)^{\eta/2a}$ and $x=-k_0^2 z^2$ in \equ{Sch} we arrive at the confluent Heun equation \cite{arXiv:0807.2219}
\begin{equation}\label{heun}
x(x-1)w''(x)+((\frac{1}{2}+\frac{\eta}{k_0})x-\frac{1}{2})w'(x)+\frac{m^2}{4k_0^2}(1-x)w(x)=0,
\end{equation}
where $\Phi(z)=w(x)$. The right massless mode  $f_0(z)=C_0(1+k_0^2z^2)^{\eta/2k_0}$, is normalized for $\eta/k_0<-1/2$. The solution of \equ{Sch} for massive right fermion is
\begin{equation}
 \psi(z)=\left(AH_1(-k_0^2z^2)+BH_2(-k_0^2z^2)\right)(1+k_0^2 z^2)^{\eta/2k_0},
\end{equation}
where $H_1(x)$, $H_2(x)$ are the confluent Heun functions\cite{arXiv:0807.2219} and $A$,$B$ are constants. In this case, we show the transmission 
coefficient in Fig.\ref{fig4} for $k_0=1$ and $\eta=-1$. With these parameters we do not have massive resonant modes.

\subsection{Discrete massive modes}

Looking at the expression of the potential, given by Eq. \equ{U(A)}, we can 
study what type of $F(\phi)$ coupling gives discrete massive modes. 
Asymptotically, for any warp 
factor we must have $A(z)\sim 1/z$. In order to get bounded states, $U(z)$ must go to a constant or to infinity. In the case of $U(z)$ asymptotically going to a constant, we must have that $F(\phi)\sim z$.  For the warp factor given by Eq. \equ{A3} and  $F(\phi)=\tan(\phi/\sqrt{3})=k_0z$, the right potential is
\begin{equation}
 U(z)=\frac{k_0\eta}{(1+k_0^2 z^2)^{3/2}}+\frac{k_0^2 \eta^2 z^2}{1+k_0^2 z^2} .
\end{equation}
\begin{figure}[ht]
 \centerline{\includegraphics[scale=0.5]{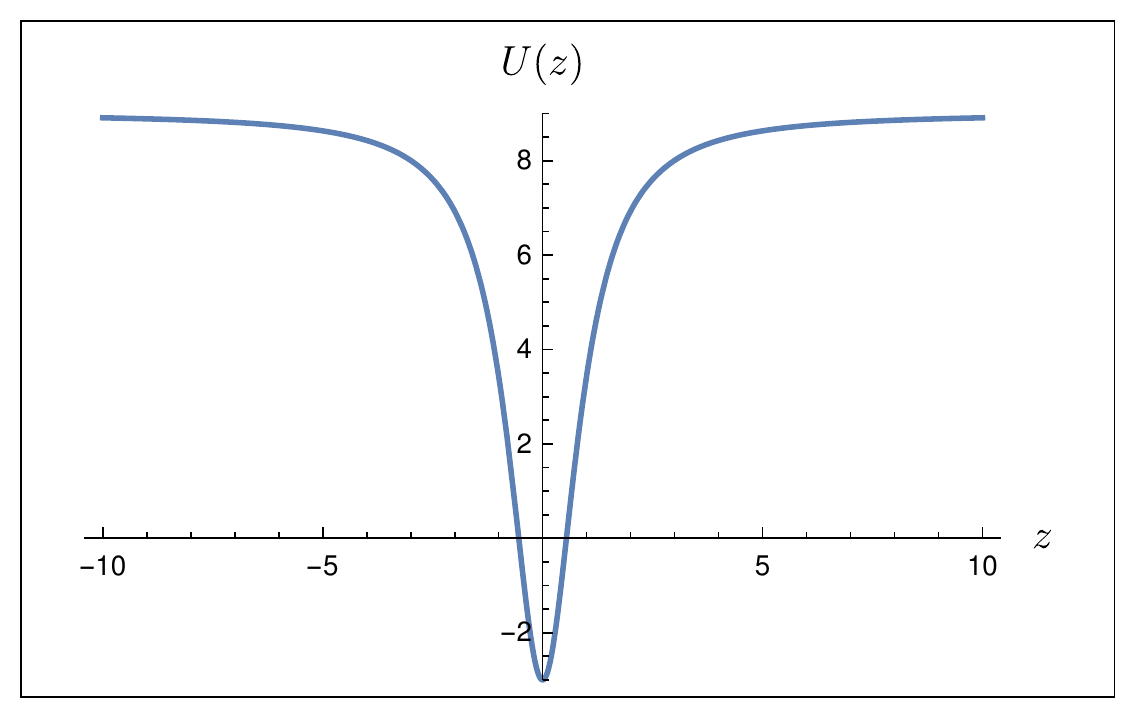}}
 \caption{The right potential profile for $F(\phi)=\tan(\phi/\sqrt{3})$, with $k_0=1, \eta=-3$.}
 \label{fig5}
\end{figure}

In this case, the right massless mode is given by $f_0(z)=C_0e^{\frac{\eta}{k_0}\sqrt{1+k_0^2z^2}}$ and is normalized for $\eta<0$.
For the potential in Fig.\ref{fig5}, we obtain  discrete massive modes 
$m=2.16,2.67,2.87$  for $k_0=1$ and $\eta=-3$. 

It can be shown that is  possible to obtain 
analytically discrete localizable modes. As mentioned before, the massless right 
mode obey $\eta e^A F(\phi)=f_0'/f_0$,
then the right and left potential reads
\begin{eqnarray}
 &&U_{+}=\frac{f_0''}{f_0},\\
 &&U_{-}=2\left(\frac{f'_0}{f_0}\right)^2-\frac{f_0''}{f_0}.
\end{eqnarray}
From the arbitrariness of $F(\phi)$, we can choose $f_0(z)$ in order to obtain normalizable modes for $\psi_{+}$ and
$\psi_{-}$. Let us consider the case where $U_+-U_-=\gamma$, with $\gamma$ constant. 
Then we obtain the solution
\begin{eqnarray}
 f_0(z)=Ce^{\frac{\gamma}{4}z^2},\\
 U_{\pm}(z)=\frac{\gamma^2}{4}z^2\pm\frac{\gamma}{2}.\\
\end{eqnarray}
Since $f_0(z)$ is the zero mode for right fermion, the localization occurs only if $\gamma<0$. In this case, taking
$\gamma=-2\omega^2$, we obtain the schrodinder-like equation
\begin{eqnarray}
 -\psi''_{\pm}(z)+(\omega^4 z^2\mp\omega^2)\psi(z)_{\pm}=m^2\psi_{\pm}(z).
\end{eqnarray}
This is the exactly the Schr\"odinger equation for the one dimensional harmonic oscillator. The massive modes are localizable
only if $m=\sqrt{2n}~\!\omega$,  for right fermions and 
$m=\sqrt{2n+2}~\!\omega$, for left fermions, $n=0,1,2\cdots$. This gives the 
coupling
\be{FPHI}
\eta F(\phi)=-\frac{\omega^2}{k_0}\frac{\sin(\phi/\sqrt{3})}{\cos^2(\phi\sqrt{3})}.
\ee

This type of coupling was studied in another context in \cite{Liu:2009dwa} 
for a symmetric $dS$ branes. This coupling provides analytical treatment and, 
in this particular case,  gives a remarkable characteristic: 
localizable massive modes. It is then important to discuss this coupling in order to find some physical interpretation, if it exists in fact.  Despite this, 
massive fermionic modes can be found in the context of inflating baby-skyrmion brane models \cite{Delsate:2011aa, Kodama:2008xm} and in other more usual 
scenarios \cite{Ringeval:2001cq}.
\section{Conclusions and perspectives}
Here we have shown that it is possible to generalize the formalism to obtain analytical solutions to models of localizations of fermions in RS scenarios. 
This work is a generalization of previous results only applied to bosonic fields, namely, the gravitational, vector and scalar fields. The main idea is to 
make use of a warp factor defined in a piecewise way and solve an associated equation that describes its profile centered at the origin. With the analytical 
solution at hand it is quite easy to obtain and study the resonances in the models. We analyzed the conditions of fermion-scalar coupling which give localizable massive modes. 
For a very specific case we found an analytical coupling between fermions and 
the scalar field  which gives localizable discrete massive modes. For a very specific case we found an analytical coupling between fermions and the scalar field  which gives
localizable discrete massive modes. These results are very interesting since the mass spectrum of particles is discrete. 
 We believe that, despite the fact we are studying fermions, this formalism could open a new way to approach open problems as analytical study of
 non-abelian gauge field in branes\cite{Alencar:2015awa}.
More interesting yet is the discussion of models containing another kind of spinors, a result which 
can be applied to supersymmetric models. All of these questions are left for future works.

\section{Acknowledgments}

We acknowledge the financial support provided by Funda\c c\~ao Cearense de Apoio ao Desenvolvimento Cient\'\i fico e Tecnol\'ogico (FUNCAP), the Conselho Nacional de 
Desenvolvimento Cient\'\i fico e Tecnol\'ogico (CNPq) and FUNCAP/CNPq/PRONEX.

%

\end{document}